\documentclass[sigconf]{acmart}
\copyrightyear{2021}
\acmYear{2021}
\setcopyright{acmcopyright}\acmConference[ESEM '21]{ACM / IEEE International Symposium on Empirical Software Engineering and Measurement (ESEM)}{October 11--15, 2021}{Bari, Italy}
\acmBooktitle{ACM / IEEE International Symposium on Empirical Software Engineering and Measurement (ESEM) (ESEM '21), October 11--15, 2021, Bari, Italy}
\acmPrice{15.00}
\acmDOI{10.1145/3475716.3475769}
\acmISBN{978-1-4503-8665-4/21/10}

\AtBeginDocument{%
  \providecommand\BibTeX{{%
    \normalfont B\kern-0.5em{\scshape i\kern-0.25em b}\kern-0.8em\TeX}}}

\usepackage{tikz}
\usepackage{subfig}
\usepackage{graphicx}
\usepackage{makecell}
\usepackage[utf8]{inputenc}
\usepackage{tcolorbox}

\usepackage{amsmath}
\usepackage{stfloats}

\usepackage{enumitem}
\newtcbox{\inlinecode}{on line, boxrule=0pt, boxsep=0pt, top=2pt, left=2pt, bottom=2pt, right=2pt, colback=gray!15, colframe=white, fontupper={\ttfamily \footnotesize}}

\begin{document}

\title{A comparative study of vulnerability reporting by software composition analysis tools}

\author{Nasif Imtiaz}
\email{simtiaz@ncsu.edu}
\affiliation{
  \institution{North Carolina State University}
  \state{North Carolina}
  \country{USA}
}

\author{Seaver Thorn}
\email{swthorn@ncsu.edu}
\affiliation{
  \institution{North Carolina State University}
  \state{North Carolina}
  \country{USA}
}

\author{Laurie Williams}
\email{lawilli3@ncsu.edu}
\affiliation{
  \institution{North Carolina State University}
  \state{North Carolina}
  \country{USA}
}

\begin{abstract}

\textbf{\textit{Background:}} 
Modern software uses many third-party libraries and frameworks as \textit{dependencies}. Known vulnerabilities in these dependencies are a potential security risk. Software composition analysis (SCA) tools, therefore,
are being increasingly adopted by practitioners to keep track of vulnerable dependencies. 
\textbf{\textit{Aim:}} The goal of this study is to understand the difference in vulnerability reporting by various SCA tools. 
Understanding if and how existing SCA tools differ in their analysis
may help security practitioners to choose the right tooling 
and identify future research needs.  
\textbf{\textit{Method:}} We present an in-depth case study 
by comparing the analysis reports of 9 industry-leading SCA tools 
on a large web application, OpenMRS, 
composed of Maven (Java) and npm (JavaScript) projects.
\textit{\textbf{Results:}} 
We find that the tools vary in their vulnerability reporting. 
The count of reported vulnerable dependencies ranges 
from 17 to 332 for Maven and 
from 32 to 239 for npm projects across the studied tools. 
Similarly, the count of unique known vulnerabilities 
reported by the tools ranges from 36 to 313 for Maven 
and from 45 to 234 for npm projects. 
Our manual analysis of the tools' results 
suggest that accuracy of the vulnerability database 
is a key differentiator for SCA tools. 
\textbf{\textit{Conclusion:}} We recommend that
practitioners should not rely on any single tool at the present, 
as that can result in missing known vulnerabilities.
We point out two research directions in the SCA space:
i) establishing frameworks and metrics to identify 
false positives for dependency vulnerabilities; 
and ii) building automation technologies for continuous monitoring 
of vulnerability data from open source package ecosystems. 

\end{abstract}


\maketitle

\section{Introduction}

Most modern software uses third-party open source libraries, packages, or frameworks that are referred to as \textit{dependencies}. 
A Black Duck report found 98\% of the 1,546 
audited commercial codebases in 2020
contained open source packages 
with an average of 528 packages per codebase~\cite{blackduck2021}.
% The rise of open source use can be evident 
% from the size of npm, a JavaScript package manager, that hosts more than 1.5 million packages at the time of this writing~\cite{nodemodules}. 
However, known vulnerabilities in dependencies 
are one of the top ten security risks~\cite{owasp29top}. 
The Black Duck audit also found 84\% of the codebases to contain at least one publicly known vulnerability in their open source dependencies~\cite{blackduck2021}.

Software composition analysis (SCA) tools 
are used to report known vulnerabilities in the open source dependencies of a software. 
However, these tools may differ in how they detect the dependencies and the vulnerability database they maintain. 
A comparative study is yet to be performed 
to review the existing SCA tools and
determine if and how they differ. 
Furthermore, not all alerts generated by SCA tools 
are relevant or high priority to the developers~\cite{pashchenko20qualitative}. 
If and how existing SCA tools aid developers in assessing the risk of the vulnerabilities from the context of the client application needs to be studied to help future research. 

The goal of this study is 
to aid security practitioners and researchers 
in understanding the vulnerability reporting 
by software composition analysis tools 
through a comparative study of these tools 
on a real-world case study. 
Our research questions are:

\begin{quote}
    \textbf{RQ1:} What are the differences between vulnerability reports produced by the different software composition analysis (SCA) tools?
\end{quote}

\begin{quote}
    \textbf{RQ2:} What metrics are presented by the SCA tools to aid in the risk assessment of dependency vulnerabilities?
\end{quote}

To answer, we present an in-depth case study 
by running 9 SCA tools 
on a large web application, OpenMRS, 
that utilizes two popular package ecosystems. 
The application consists of 43 Maven (Java) 
and 5 npm (JavaScript) projects.  
The studied SCA tools vary in their scanning technique 
and vulnerability database, 
and represent the state-of-the-art. 
% We have included both open source and commercial tools based on their popularity and availability.
The contributions of this paper include 
the first evaluation of the SCA tools through 
(a) a quantitative comparison of their vulnerability reports 
on a real-world case study, 
(b) a manual analysis of the differences 
among the tools' reports, 
and (c) characterization of metrics provided by the tools for assessment of the dependency vulnerabilities. 

The remainder of the paper is structured as follows: 
Section \ref{terminologies} introduces
the key concepts and terminologies; 
Section \ref{sec:openmrs} and \ref{sec:tools} explains the evaluation case study and the studied SCA tools.  
Section \ref{sec:findings} discusses the findings of this paper, followed by discussion and limitations of the findings. 
Section \ref{sec:relwork} discusses related work, followed by conclusion.

\section{Key Concepts \& Terminologies:} \label{terminologies}
\textbf{Dependency:} When a software uses an open source package, the package is referred to as \textit{\textbf{dependency}} of the software. Typically, a software declares a specific or a range of valid versions of a package as its dependency in a manifest file that we refer to as \textit{\textbf{dependency file}}. However, a software may use open source package or code fragments without explicit declaration as well~\cite{scaevaluation}. In the remainder of this paper, we refer to `dependency' as a specific version of a package. For example, version 1.0.0 and version 2.0.0 of the same package \textit{A} will be considered as distinct \textit{\textbf{dependencies}}. However, they will be considered as the same \textit{\textbf{package}}.

The dependencies declared through dependency files are resolved through some \textit{\textbf{package manager}}. \textit{pom.xml} and \textit{package.json} are dependency files for Maven and npm package manager, respectively. The dependencies that a software accesses
directly from its own code are called \textit{\textbf{direct}} dependencies. However, the direct dependencies may depend on other open source packages that are required by the host machine
to run the software successfully. Such packages are called \textit{\textbf{transitive}} dependencies. Therefore, for most package managers, including Maven and npm, the whole dependency structure is hierarchical and forms a tree format. The \textit{\textbf{depth}} of a dependency refers to their level 
in the dependency tree, 
with direct dependencies having a depth of one.

\textbf{Vulnerability:} \label{bg:vulnerabilities}
NIST~\cite{NISTDefinition} defines vulnerability  as 
``weakness in an information system, system security procedures, internal controls,  or implementation that could be exploited or triggered by a threat source.'' If a vulnerability gets exploited by a threat source, the potential for loss or damage is referred to as \textit{\textbf{risk}} of the vulnerability. Vulnerabilities can get discovered in already released versions of software packages. If reported, respective package maintainers can fix the vulnerability in a new version. When a dependency of a software is subject to publicly known vulnerabilities, it is referred to as a \textit{\textbf{vulnerable dependency}}.

\textbf{Software Composition Analysis (SCA):} SCA is a part of application analysis that deals with managing open source use. SCA tools typically generate an inventory of all the open source components in a software product and analyze the license compliance and the presence of any known vulnerabilities in them. By the vulnerability detection capability of SCA tools, we mean the ability to identify and report known vulnerabilities in the open source components used by a software application. 

\textbf{Disclosed and Discovered Vulnerabilities:} \label{bg:cve}
The National Vulnerability Database (NVD)~\cite{nvd} is the U.S. government repository of publicly accessible standards-based vulnerability management data.  The primary reference-tracking system for publicly disclosed vulnerabilities in the NVD database is the Common Vulnerabilities and Exposure (CVE) system where each vulnerability is referenced by a unique \textbf{CVE} identifier, a system developed by Mitre~\cite{mitre}.   

Additionally, SCA tools augment NVD vulnerabilities/CVEs with vulnerabilities found in other databases, such as npm Security Advisories~\cite{nodesecurityio}, Sonatype OSS Index~\cite{ossindex}, and GitHub security advisories~\cite{ghsa}, that do not necessarily have a CVE identifier. Similarly, SCA tools can also have proprietary techniques to discover vulnerabilities in open source packages~\cite{snykinfo, zhou2017automated} as explained in Section \ref{sec:selection}. In this paper, vulnerabilities reported by SCA tools that do not have an associated CVE identifier are referred to as \textit{\textbf{Non-CVEs}}.

\subsection{Maven:}
Maven is a package manager
for Java projects.

\textbf{Dependency Scopes:} Maven dependencies can have six different scopes~\cite{Mavendep}: compile, provided, runtime, test, system, and import. The scopes determine the phase when a dependency will be used and if the dependencies can propagate transitively.

% Below are the different dependency scopes 
% under Maven\cite{Mavendep}:

% \begin{enumerate}
%     \item Compile: Compile dependencies are available in all classpaths of a Java project and propagates to dependent packages as transitive dependencies. This is the default scope in Maven.
%     \item Provided: A dependency that is required during compilation and testing but is not added to the Java project's runtime classpaths. Therefore, the user has to provide this dependency in runtime. Provided dependencies do not propagate as transitive dependencies.
%     \item Runtime: This scope indicates that the dependency is not required for compilation, but is required for execution. Maven includes dependency with this scope in the runtime and test classpaths, but not compile classpaths. Runtime dependencies propagate as transitive dependencies.
%     \item Test: The dependency with this scope is not required for normal use of the software, and is only required in the testing phases. The scope is not transitive.
%     \item System: This scope is similar to provided except that user has to explicitly provide the dependency jar file (not automatically managed by the package manager).
% \end{enumerate}

\textbf{Dependency Mediation:} 
When there are multiple versions of a package in the dependency tree, Maven picks one with the nearest definition.
% Maven picks the nearest definition -- that is --
% the version with the smallest depth
% and the one that is first declared among equal depths.
Therefore, usually, a single project has a single version of a package as a dependency
that is read from a local repository. In the dependency file (\textit{pom.xml}), developers generally specify a single version for its dependencies. Version numbers can have up to five parts indicating major, minor, or incremental changes.

% Version numbers for Maven packages 
% can consist of five parts~\footnote{\url{https://docs.oracle.com/middleware/1212/core/Maven/Maven_version.htm#Maven400}}: a) major version, b) minor version, c) incremental version, d) build numb, e) qualifier. 
% While explanation of each part is out of scope of this paper,
% a change in major version indicates major changes and may affect the communication between a dependency and a dependent package.

\subsection{Node Package Manager (npm):}

npm is a package manager for JavaScript projects.

\textbf{Dependency Scopes:} npm has two primary dependency scopes: \textit{Prod} (production) and \textit{dev} (development) to indicate the phase where a dependency is required. 
% Two other dependency scopes, 
% peer and optional, are not automatically resolved by npm and therefore, 
% are ignored in this study.

% Below are the different dependency scopes 
% under npm:
% \begin{enumerate}
%     \item Production (Prod): Packages required by the software application in production.
%     \item Development (Dev): Packages that are only required for local development and testing.
%     \item Peer: Peer dependencies are not automatically installed by the package manager as transitive dependencies,
%     but may require to be installed by the dependent software as a \textit{peer} for some functionality. Typically, when the dependent application and a dependency both uses another package and requires a some for of communication among them, the user needs to install the package explicitly as a direct dependency to ensure same versioning and dependency path.
%     \item Optional: The dependency is not required to run its core functionality and not automatically installed by the package manager.
% \end{enumerate}

\textbf{Dependency Mediation:} npm copies all the dependencies in a project subdirectory called `node\_modules', with a similar structure of the dependency tree. If two dependencies \textit{A} and \textit{B} both depend on the same package \textit{C}, two different copies of package \textit{C} will be copied inside package \textit{A} and \textit{B}. Therefore, the same dependency can have multiple paths to be introduced to the root application. Also, the same package can have multiple versions
as dependencies. Therefore, npm has a concept called \textit{\textbf{dependency path}}, which is not present in Maven. Each unique path a dependency is introduced to the root application is referred to as the \textit{dependency path}. In npm, developers can list a range of versions for a package that is valid as a dependency. npm also has the concept of \textit{\textbf{lock}} files -- a snapshot of the entire dependency tree and their resolved version at a given time; and can be used to instruct npm to install the specified versions in the lock file. npm packages follow the SemVer format~\cite{semver} for version numbering.

% \textbf{Version numbering:} npm packages follow the SemVer format~\cite{semver} for version numbering which has three parts indicating
% a) major release (may break backward compatibility), 
% b) minor release (backward compatible new features), 
% c) patch release (backward compatible bug fixes).

\section{Evaluation Case Study: OpenMRS}\label{sec:openmrs}
OpenMRS is a web application for electronic medical record platform~\cite{openmrs}. A particular configuration of OpenMRS that can be installed and upgraded as a unit is referred to as a \textit{distribution}. The general purpose distribution of OpenMRS is the ``Reference Application Distribution''~\cite{referenceapplication}.
We choose Version 2.10.0 of this distribution released on April 6, 2020 (the latest release at the time of this study
% ; 1,168,351 lines of code
) as our evaluation subject. In the remainder of the paper, we refer to the whole distribution simply as ``OpenMRS''.

OpenMRS consists of 44 projects that are hosted in their own separate repositories on GitHub. 
% We consider each project hosted on its own repository as an individual entity. 
Out of the 44 projects, 39 are Maven projects and 1 is a npm project. The other 4 projects are composed of a Maven and a npm project each.
% We identify Maven and npm project based on the presence of \textit{pom.xml} and \textit{package.json} files that specify metadata for corresponding projects. 
Based on OpenMRS structure, \textbf{we scope our study to Maven and npm dependencies.}
We use OpenMRS SDK~\cite{openmrssdk} to automate the build, test, and run of the individual projects and assemble the full application in this study.

\subsection{Why OpenMRS?}\label{sec:whyopenmrs}
Choosing test cases to evaluate software security tools 
can be a complex task.
For comparison of security tools, 
Delaitre et al.~\cite{delaitre2018sate} notes
that the test case should have \textit{sufficient and diverse number} of security weaknesses.
OpenMRS depends on many third-party dependencies 
as will be seen in Section \ref{sec:openmrsdeps}; 
and being a web application, 
is composed of several heterogeneous components,
such as database, content generation engines,
client-side code etc.,
therefore increasing the probability of having 
a large, diverse set of vulnerable dependencies.

Another approach of comparison 
instead of a single case study 
can be running the tools on a group of diverse projects. 
However, three of the selected tools in this study (Steady, Commercial A, and B) are (a) resource and time-consuming 
to set up and run; 
(b) involve certain requirements, e.g., 
acceptance tests for interactive binary instrumentation, 
unit tests for executability tracing; 
and (c) involve permission issues in case of the commercial tools. On the contrary, focusing on a single case study enables us to manually investigate the differences in the tools' results.

OpenMRS has also been used in security research in the past~\cite{crain2017open, tondel2019collaborative,de2012overview, lamp2018danger,rizvi2015relationship,amir2016correct}. ~\cite{lamp2018danger} evaluated OpenMRS for medical system security requirements; ~\cite{rizvi2015relationship} evaluated OpenMRS for access control checking; while ~\cite{amir2016correct} studied OpenMRS for correct audit logging.

\subsection{OpenMRS: Dependency Overview}\label{sec:openmrsdeps}
In this section, we provide an overview of Maven and npm dependencies of OpenMRS. We parse the dependency tree of each project through native \inlinecode{mvn dependency:tree} and \inlinecode{npm list} command. We also parse each dependency's scope and depth in the dependency tree.

Table \ref{tab:depoverview} provides a dependency overview of OpenMRS. Note that, for Maven projects, there can be \textit{internal} dependencies -- that is -- a project within the OpenMRS distribution can be listed as a dependency for another project. We do not count the internal dependencies in Table \ref{tab:depoverview}. Also, npm projects can contain \textit{lock} files such as \textit{shrinkwrap.json}, \textit{package-lock.json} which are not considered.

\begin{table}[]
    \centering
    \caption{OpenMRS dependency overview}
    \label{tab:depoverview}
    \begin{tabular}{lrr}
         &  \textbf{Maven} & \textbf{npm} \\
         \hline
         No. of projects & 43 & 5  \\
         \hline
         \makecell[l]{Total unique dependencies\\(package and version)} & 547 & 2,213\\
         Total unique packages & 311 & 1,498 \\
         \hline
         Median dependency per project  & 127.0 & 840.5\\
         Median dependency path per project & NA & 1,675.0 \\
         \hline
         Median depth of dependencies & 2 & 4\\
         Max. depth of dependencies & 7 & 12 \\
         \hline
         Median Provided dependencies & 99.0 & NA\\
         Median Compile dependencies & 3.0 & NA\\
         Median Runtime dependencies & 5.0 & NA \\
         Median Test dependencies & 24.5 & NA\\
         \hline
         Median Production dependencies & NA & 202.5 \\
         Median Production dependency path & NA & 366.0 \\
         Median Developer dependencies & NA & 807.5 \\
         Median Developer dependency path & NA & 1,613.5 \\
         \hline
         
    \end{tabular}
\end{table}

\section{SCA Tools}\label{sec:tools}
In this section, we explain the criteria we use to select the SCA tools; description of the tools; how we performed the scan on OpenMRS; and how we analyzed the reports produced by the tools.

\subsection{Selection Criteria}\label{sec:selection}
To identify the existing SCA tools from both industrial offerings and the latest research, we performed an academic literature search and a web search through the following keywords:
(\textit{vulnerable} OR \textit{open source} OR \textit{software}) AND (\textit{dependency} OR \textit{package} OR \textit{library} OR \textit{component} OR \textit{composition}) AND (\textit{detection} OR \textit{scan} OR \textit{tool} OR \textit{analysis}). From the relevant search results, we filtered the tools with the following \textit{inclusion criteria}:
a) scans either Maven or npm projects; 
b) we have access to an executable tool; 
and c) offers unique features when compared 
with already selected tools. 
From our selection process, we selected nine tools. 
Two of the tools are not freely available, 
and the license agreements prevent us from providing names. 
We refer to them as Commercial A and B. \textbf{Out of the selected 9 tools, 4 tools can scan both Maven and npm projects, 1 tool scans only npm projects, while 4 tools scan only Maven projects. }

We observed that SCA tools primarily differ in three dimensions:
\begin{enumerate}
    \item \textbf{Vulnerability database:} 
    To report the list of known vulnerabilities, the tools need a database. Tools can pull vulnerability data from \textit{third-party} source(s) such as NVD CVEs~\cite{nvd}.
    Additionally, SCA tools can maintain their own vulnerability database where they collect and verify vulnerability data through different techniques~\cite{snykinfo, zhou2017automated}.
    % Additionally, SCA tools can maintain their own vulnerability database where they may a) verify vulnerabilities and their metadata (e.g. what versions of what packages are affected) from other databases, b) enhance the description and severity ratings of the vulnerabilities, c) monitor and triage available exploits, d) discover vulnerabilities in open source packages through proprietary techniques (e.g. identifying vulnerabilities possibly discussed in various forums yet not officially disclosed yet)~\cite{snykinfo, zhou2017automated}.
    \item \textbf{Dependency scanning source:} SCA tools can detect open source dependencies from dependency manifest file, source code, and binaries. Typically, dependency files are the common source to resolve dependencies of a project, as is done by the package managers as well.
    \item \textbf{Additional analysis to infer dependency use:}  Tools can perform additional static and/or dynamic analysis to infer how the dependencies are being used by an application.
\end{enumerate}

\subsection{Tool description} \label{sec:runningtools}
For the selected tools, we describe 
(a) if they scan Maven or npm dependencies, 
their (b) data source, (c) scanning technique, 
and (d) how we performed the scan for this study.

% \begin{enumerate}

\textbf{OWASP Dependency-Check (DC):} This tool scans 
both Maven and npm projects and works by scanning the dependency files, JARs, and JavaScript files~\cite{owaspdc}. It pulls vulnerability data from multiple third-party sources including NVD, OSS Index, npm advisories. We used the Maven plugin to scan Maven projects and the command line tool (Version 5.3.2) to scan npm projects. We had the \textit{experimental analyzer} option enabled  to perform JavaScript scanning.

\textbf{Snyk:}
Snyk also scans both Maven and npm projects. The tool works by scanning dependency files~\cite{snyk} and maintain its own vulnerability database~\cite{snykdb}. We ran the command line tool (Version 1.382.0) that is freely available through the command \textit{snyk test --all-projects --dev --json}. 
% The \inlinecode{-dev} flag ensures that \textit{dev} dependencies will also get scanned in case of npm projects. 
% The tool produces output that shows CVSS scores both for CVEs and the non-CVEs and provides a severity level based on that.~\footnote{\url{https://support.snyk.io/hc/en-us/articles/360001040078-How-is-a-vulnerability-s-severity-determined-}}.
% Although critical is still counted as high.

\textbf{GitHub Dependabot:}
Dependabot scans both Maven and npm projects hosted on GitHub. GitHub maintains its own vulnerability database~\cite{ghsa} where it pulls data from NVD, npm advisories. Additionally, maintainers on GitHub can publish vulnerabilities in their projects as well. We hosted the 44 studied projects on the first author's GitHub account and retrieved the Dependabot alerts through GitHub API. 

\textbf{Maven Security Versions (MSV):} This tool only scans Maven projects~\cite{victims} through dependency files. We ran this tool through its Maven plugin.

\textbf{npm audit:}
This is a native tool of npm package manager for scanning npm projects. The tool works by scanning dependency files and maintains its own vulnerability database~\cite{npmadvisory}. We used the \inlinecode{npm audit --json} command.
% and the \inlinecode{npm audit --fix --dry-run --json}

 \textbf{Eclipse Steady:}
This tool only scans Java (Maven) projects. The tool performs additional analysis to assess the execution of vulnerable code in the dependencies of an application~\cite{steady}. The approach implemented is described in ~\cite{ponta2018beyond} and ~\cite{plate2015impact}. The tool requires a manual set up, 
along with the vulnerability database provided by the tool. 
We used Version 3.1.10 of this tool. 
We set up Steady in a virtual machine, allocating 16 GB RAM,
and 4 processor cores. Steady hosts their vulnerability data set on GitHub~\cite{steadydata}. The data set contains patch commit information for each vulnerability. We imported the data source updated on Jan 24, 2020. We then performed the patch analysis feature provided by the tool to identify the involved code constructs for each vulnerability. 
For reachability analysis of the identified vulnerabilities, Steady performs three analyses: 1) static call graph construction; 2) executing JUnit tests for analyzing executability traces; and 3) JVM instrumentation through integration testing. We were unable to complete the third analysis
as the tool presumably ran out of memory after running for ten days.
% The tool execeeded 
% We performed the first two analysis.
% The binary instrumentation for the third analysis
% was still running for the eighth day by the tool we set up.
% Therefore, we were unable to successfully complete the third analysis.
% and provide results for it.

\textbf{Commercial A:} This tool has scientific papers discussing their approach (not citing to maintain blindness). We contacted their research team and provided them with the repository links for the studied projects. They returned to us with scan reports only for Maven dependencies for 37 projects and reported that they failed to complete the automated scans for the rest of the projects which may have required manual intervention. This tool offers static analysis by default and dynamic analysis as an option to identify vulnerable call chains. We received results only with static analysis performed on the code. The tool maintains its own vulnerability database. 
% Besides NVD, the tool monitors software and bug repositories to discover new open source vulnerabilities through a semi-automated method. 

 \textbf{Commercial B:} We used the free cloud edition of the tool, where it only scanned the Java dependencies (the customer support informed us that the tool does not scan front-end libraries). The tool checks for the reachability of vulnerabilities in dependency through interactive application security testing -- that is -- monitoring dependencies in use when an application is run and interacted with either through automated testing or human testers. The tool uses third-party vulnerability databases including NVD, which they curate themselves to enhance accuracy. 
%  Furthermore, the tool has their own vulnerability discovery process in open source. 
To run this tool on OpenMRS, we make use of 123 test cases provided by OpenMRS for integration testing that interact with the application through a Selenium web-driver. We connected OpenMRS to this tool and used the integration test suite to interact with the application.

\textbf{Commercial C:} This tool has a free/limited offering as
 a third-party application (bot) on GitHub, which scans both Maven and npm projects. Com. C also maintains its own vulnerability database. We connected the GitHub bot with our hosted repositories and retrieved the issues created by the bot through GitHub API.

We collected the scan reports separately for 44 projects for 8 tools. For Commercial B, which analyzes the application during runtime, 
we get a single report for the whole OpenMRS distribution. 
As vulnerability data gets updated over time, 
we ran all the tools during September 2020 
to ensure a fair comparison, except for Steady 
whose vulnerability data is from January 2020.

\begin{table*}[]
    \centering
    \caption{Vulnerable Dependencies for Maven (Java) projects}
    \label{tab:javaalerts}
    \begin{tabular}{lrrrrrrr}
        \textbf{Tool} & \textbf{Alert} & \makecell{\textbf{Unique}\\\textbf{Dependency}} & \makecell{\textbf{Unique}\\\textbf{Package}} & \makecell{\textbf{Unique}\\\textbf{Vulnerability}} & \textbf{CVE} & \textbf{Non-CVE} & \makecell{\textbf{Scan Time}\\ \textbf{(Minutes)}}   \\
        \cline{2-5}
         & \multicolumn{4}{c}{\textbf{Total (Median per project)}} \\
         \hline
         OWASP DC & 12,466 (254.0) & 332 (38.0) & 149 (36.0) & 313 (117.0) & 289 &  24 & 14.4 \\
         Snyk & 4,902 (66.0) & 96 (6.0) & 46 (6.0) & 189 (23.0) & 178 & 11 & 15.1\\
         Dependabot & 136 (0.0) & 20 (0.0) & 11 (0.0) & 61 (0.0) & 61 & 0 & NA\\
         MSV & 3,197 (58.0) &  36 (12.0) & 14 (12.0) & 36 (22.0) & 36 & 0 & 3.4\\
         Steady & 2,489 (51.0) & 91 (20.0) & 39 (19.0) & 97 (41.0) & 89 & 8 & 385.0 \\
         Commercial C & 434 (0.0) & 76 (0.0) & 44 (0.0) & 146 (0.0) & 127 & 19 & NA\\
         Commercial A & 2,998 (70.0) & 107 (24.0) & 53 (24.0) & 208 (70.0) & 187 & 21 & NA   \\
         Commercial B & 205 & 35 & 35 & 127 & 127 & 0 & NA \\
        %  Commercial C & 57 & 17 & 17 & 57 & 57 & 0 & NA \\
    \end{tabular}
    
\end{table*}
\begin{table*}[]
    \centering
    \caption{Vulnerable Dependencies for npm (JavaScript) projects}
    \label{tab:javascriptalerts}
    \begin{tabular}{lrrrrrrrr}
        \textbf{Tool} & \textbf{Alert} & \makecell{\textbf{Unique}\\\textbf{Dependency}\\\textbf{Path}}  & \makecell{\textbf{Unique}\\\textbf{Dependency}} & \makecell{\textbf{Unique}\\\textbf{Package}} & \makecell{\textbf{Unique}\\\textbf{Vulnerability}} &
        \textbf{CVE} & \textbf{\makecell{Non-\\CVE}} & \makecell{\textbf{Scan Time}\\ \textbf{(Minutes)}}  \\
        \cline{2-6}
         & \multicolumn{5}{c}{\textbf{Total (Median per project)}} \\
         \hline
         \\
         \makecell[l]{OWASP DC} & 1,379 (208.0) & 498 (72.0) & 239 (71.0) & 160 (57.0) & 234 (71.0) & 78 & 156 & 4.4\\
         Snyk & 2,210 (135.0) & 1,004 (44.0) & 90 (20.0)  & 54 (17.0) & 121 (26.0) & 79 & 42 & 1.0\\
         \makecell[l]{Dependabot} & 97 (8.0) & NA & 32 (1.0) & 30 (1.0) & 45 (4.0) & 29 & 16  & NA\\
         npm audit & 1,266 (37.0) & 852 (28.0)  & 58 (12.0) & 45 (12.0) & 62 (16.0) & 31 & 31 & 0.1  \\
         Commercial C & 205 (32.0) & 205 (32.0) & 89 (14.0) & 55 (9.0) & 96 (18.0) & 58 & 38 & NA \\
    \end{tabular}
\end{table*}

\subsection{Analyzing Tool Results}\label{sec:analysismethod}
Below, we discuss the metrics and information
that we processed from the tool reports 
to answer our research questions.

\textbf{Quantity of Alerts:} 
When a project is scanned by a tool, the tool reports a raw count of alerts identified on the project. However, the alerts do not represent either unique dependencies or unique vulnerabilities. We observed that the same alerts can be repeated in tools' reports for various reasons. The alert count, however, 
may indicate the amount of audit effort required from the developers.
% The same alerts can be repeated due to
% OpenMRS's multi-module project structure.

\textbf{Tracking unique dependency, dependency path, package, and vulnerability:} The definitions of these four metrics, as used in this study, are provided in Section \ref{terminologies}. 
When processing the analysis reports from all the tools, 
we store the data in a relational database schema. 
In the schema, we keep an identifier for 
each unique package, 
dependency (package:version), 
dependency path, 
and CVE identifier. 
For the non-CVEs, 
all tools except OWASP DC and Commercial A 
provide a tool-specific identifier. 
While OWASP DC and Commercial A 
provide no reliable identifier to track unique non-CVEs, 
upon manual inspection, we noticed that 
the vulnerability description along with the affected package(s) 
are a reliable way to track non-CVEs. 
However, we have no reliable way to map non-CVEs 
across different tool reports.

\textbf{Scan time} indicates the total number of minutes 
a tool took to scan all the projects. 
We have no scan time for GitHub and Com. A, 
as they are GitHub cloud services. 
We collected the issues and alerts from GitHub at the end of September, at least two weeks after hosting the repositories. Commercial B monitors dependency during runtime through interaction, therefore, also does not have a definite scan time. 

\textbf{Other information:} Tools had additional information in their reports, generally to aid developers in assessing the risk of the alerts and to help in fixing them. We also collected these additional data, which will be explained in Section \ref{sec:findings} when discussing the findings.

\textbf{Manual analysis of the tools' report:} 
To understand why there are differences in the tools' results,
we manually inspected the tools' results.
We specifically focused on the project \textit{coreapps} 
as this is the project with the largest dependency count 
and includes both Maven and npm dependencies.
The first author went through results 
from all the tools for \textit{coreapps},
and categorized the differences.
The second author then independently went through
the results from the studied tools
and verified the categorization done by the first author.

\begin{figure*}
    \centering
    \subfloat[Overlap ratios for Maven vulnerable dependencies\label{heatmapa}] 
    {{\includegraphics[scale=0.5]{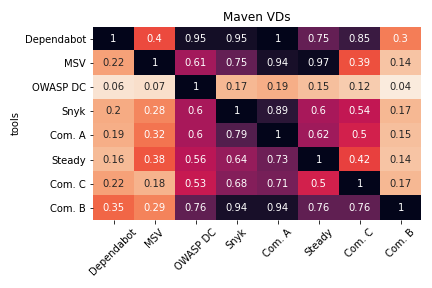}}}
    \subfloat[Overlap ratios for npm vulnerable dependencies\label{heatmapb}] 
    {{\includegraphics[scale=0.5]{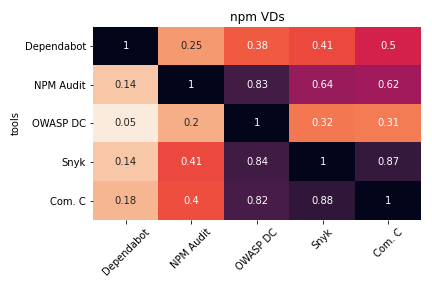}}}
    \caption{Overlap analysis of unique vulnerable dependencies for each tool pair: Cell(i, j) indicates the percentage of i'th tool's reported vulnerable dependencies that are also reported by the j'th tool.}
    \label{fig:overlapheatmap}
\end{figure*}

\begin{figure*}[h!]\label{venn}
    \centering
    \subfloat[\label{venna}]{} 
    {{\includegraphics[scale=0.275]{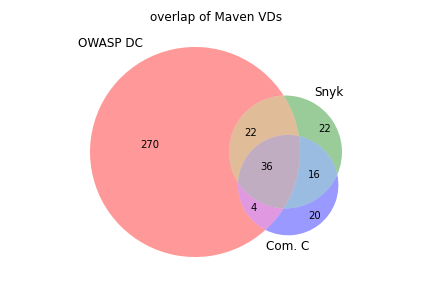} }}
    \subfloat[\label{vennb}]{}
    {{\includegraphics[scale=0.275]{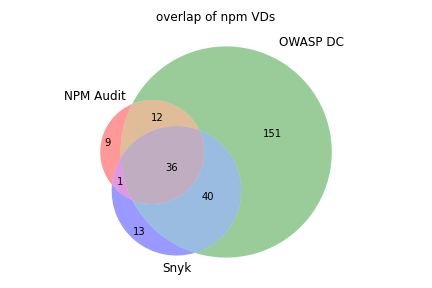} }}
    \subfloat[\label{vennc}]{}
    {{\includegraphics[scale=0.275]{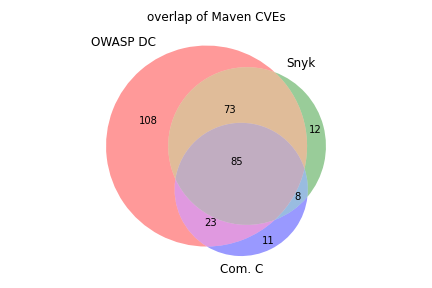} }}
    \subfloat[\label{vennd}]{}
    {{\includegraphics[scale=0.275]{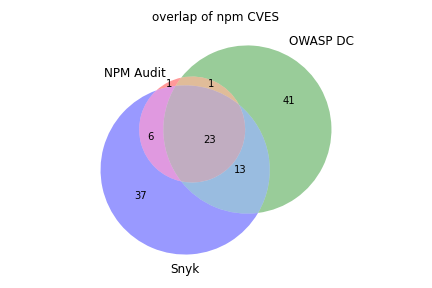} }}
    \caption{Venn Diagram for overlap of vulnerable dependencies and CVEs among three representative tools: OWASP DC, Snyk, and Com. C (Maven)/NPM Audit (npm). The sub-figures represent overlap of (a) Maven vulnerable dependencies, (b) npm vulnerable dependencies, (c) maven CVEs, (d) npm CVEs.}
    \label{fig:overlapvenn}
\end{figure*}

\section{Findings}\label{sec:findings}

In this section, we present descriptive statistics on how SCA tools differed on vulnerability detection, a manual analysis on why the tools differed (RQ1); and a characterization of the metrics provided by the studied tools for aiding in risk assessment of vulnerability in dependencies, (RQ2). 

\subsection{RQ1: What are the differences between vulnerability reports produced by the different software composition analysis (SCA) tools?}

Table \ref{tab:javaalerts} and \ref{tab:javascriptalerts} show the tools' result summary for Maven and npm dependencies, respectively. 
% Note that, we only have partial scan reports (37 projects) for Commercial A. 
The table provides the total count for alerts and unique dependencies, dependency paths (for JavaScript), packages, and vulnerabilities for OpenMRS as a single application. The tables also report the total count of CVEs, non-CVEs, and scan time.  For the eight tools that scanned projects individually (besides Commercial B), the tables provide in parentheses the median count per project for alerts and for unique dependencies, dependency paths, packages, and vulnerabilities. 

The alert counts are higher than the count of unique vulnerabilities or dependency paths, as discussed in Section \ref{sec:analysismethod}. While the total alert count repeats the same vulnerabilities found across projects, some tools repeat the same alert within a project as well due to modular project structure. We also see the unique dependency count is higher than the unique package count. Different versions of the same package may be declared as a dependency in different projects, while npm can have multiple versions of the same package as dependencies even within a single project. 
We now discuss how the SCA tools have differed in their reporting:
\begin{table*}[]
    \centering
    \caption{Scope breakdown, rate of direct dependencies among reported vulnerable dependencies (VDs), and max. depth for the reported transitive VDs}
    \label{tab:scopedepthbreakdown}
    \begin{tabular}{lrrrr|rr|rr|rr}
        \textbf{Tool} & \multicolumn{4}{c|}{\textbf{\makecell[c]{Scope breakdown for \\ Maven VDs}}} & \multicolumn{2}{c|}{\textbf{\makecell[c]{Scope breakdown for \\ npm VDs}}} & \multicolumn{2}{c|}{\textbf{\makecell{Direct VDs\\(across all projects)}}}& \multicolumn{2}{c}{\textbf{\makecell[c]{Max. Depth of\\ VDs}}} \\
         & \textbf{Compile} & \textbf{Provided} & \textbf{Runtime} & \textbf{Test} & \textbf{Prod} & \textbf{Dev} & \textbf{Maven} & \textbf{npm}  &\textbf{Maven} & \textbf{npm} \\
         \hline
         
         \makecell[l]{OWASP DC} & 58 & 66 & 4 & 54 & 65 & 207 & 5.8\% & 4.4\% & 6 & 10\\
        Snyk & 56 & 62 & 2 & 25 & 13 & 83 & 14.8\% & 3.5\% & 7 & 10\\
        \makecell[l]{Dependabot} & 15 & 5 & 1 & 2 & 6 & 8 & 97.0\% & 65.9\% & 2 & 6\\
        \makecell[l]{MSV} & 19 & 30 & 1 & 3 & NA & NA & 1.7\% & NA & 5 & NA\\
        npm audit & NA&NA&NA&NA& 15 & 51 & NA & 0.9\% & NA & 10\\
        Steady & 60 & 60 & 4 & 11 & NA & NA & 5.3\% & NA & 5 & NA \\
        Commercial C & 54 & 0 & 2 & 0 & 12 & 76 & 60.0\% & 8.8\% & 5 & 10 \\
        Commercial A & 72 & 79 & 1 & 0 & NA & NA & 8.7\% & NA & 5 & NA\\
        \hline
    \end{tabular}
\end{table*}

\textbf{The tools differed both on identifying unique vulnerable dependencies and the unique vulnerabilities:}  OWASP DC detects the highest number of unique dependencies and unique vulnerabilities for both Maven and npm projects. However, our analysis in Section \ref{sec:manual} indicates more may not necessarily be better. Conversely, Commercial B that monitors the dependency under use during runtime detected the lowest number of vulnerable dependencies for Maven projects. MSV and Dependabot detected the lowest number of unique vulnerabilities, respectively, for Maven and npm projects.

\textbf{5 out of the 8 tools for Maven reported non-CVEs while all the 5 tools for npm reported non-CVEs.} We observe that npm packages have a higher proportion of non-CVEs to CVEs than Maven packages. We find OWASP DC to report higher non-CVEs than any of the other 4 tools for npm projects. However, as OWASP DC does not provide an identifier for non-CVEs, we tracked unique non-CVEs through vulnerability description and affected packages, which may have resulted in duplication of the same vulnerabilities.

\textbf{Only 2 out of the 5 tools that scanned npm projects report vulnerable dependency paths:} 
In npm, the same package \textit{A} can be introduced transitively through multiple direct dependencies and therefore, can lie in multiple dependency paths. The developers may need to fix each path separately if there is a vulnerability in package \textit{A}. We find that only npm audit and Snyk report all possible dependency paths to each unique vulnerability. 
% For each dependency with a vulnerability in npm projects, npm audit finds a median of 2 dependency paths ($max=309$ for \textit{lodash:4.6.1} in one project). 
% Com. C on GitHub only reports one out of many possible dependency paths while OWASP DC also does not report all the possible paths. Dependabot does not report dependency path in its alerts.

% Further, npm audit provides automated fix options for each dependency path separately where possible. 

% Table \ref{tab:scopedepthbreakdown} shows a scope
% breakdown and maximum depth of identified vulnerable dependencies for 8 tools. Commercial B scans dependencies during runtime, and therefore, we are unable to determine the scope and depth for them. In the table, the same dependency can be listed under different scope for different projects and is counted twice in such cases. We find that all tools report dependencies across all scopes. Also, all tools except Dependabot, scan both direct and transitive dependencies. While maximum depth of identified vulnerable dependency for Dependabot is 2 and 6 for Maven and npm which indicates transitivity, we find that these transitive dependencies were identified

\textbf{Tools have non-overlap in reported vulnerabilities and dependencies:} We measured how much of the unique vulnerable dependencies reported by the tools overlap with each other. The heat maps in Figure \ref{fig:overlapheatmap} show overlap ratio across tool pairs for both Maven and npm projects. For a tool pair (\textit{A},\textit{B}), the heat map shows how many dependencies reported by \textit{A} were also reported by \textit{B} and vice versa. For example, for maven projects, 54\% of Snyk's reported vulnerable dependencies were also reported by Com. C. Conversely, 68\% of Com. C's reported maven dependencies were also reported by Snyk. Figure \ref{venna}, \ref{vennb} demonstrates non-overlap in dependencies through a Venn diagram for three representative tools. We can not show such a heat map for all unique vulnerabilities, as we were unable to cross-reference non-CVEs across tools. However, we also found non-overlap over reported CVEs as well across tools. Figure \ref{vennc}, \ref{vennd} demonstrates non-overlap in CVEs for OWASP DC, Snyk, and Com. C. 

\textbf{Tools detected vulnerable dependencies across all scopes and depths:} Table \ref{tab:scopedepthbreakdown} shows a breakdown of scan results per dependency scope and what portion of the reported vulnerable dependencies are introduced directly by OpenMRS. We find that reported vulnerabilities are mostly introduced through transitive dependencies, except for Dependabot and Com. C. The latter two tools assist GitHub projects in automatically fixing the vulnerable dependencies (by upgrading to a safer version), which may explain the high rate of direct dependencies in their reporting.

% To understand why the non-CVEs might not have
% been incorporated into the CVE database,
% we look at their publish date.
% For the 53 non-CVEs reported by Snyk,
% 41 were published before 2020;
% while for 54 non-CVEs reported by Com. C,
% 50 were published before 2020.
% Therefore, developers may question why
% a reported vulnerability does not have a CVE identifier
% as CVE validation usually takes around three months
% and the scan results are from September 2020.
% For npm audit, the publish date was 
% usually not present in the results.
% However, if valid,
% the presence of non-CVEs can be an indicator of
% the richness of a tool's vulnerability database.

% \textbf{Projects are often scanned in less than a minute:}
% We find tools provide VD reports 
% usually under less than a minute per project.
% The short scan time can make it possible
% to integrate VD analysis in continuous integration (CI) tools.
% GitHub can present its alerts at each code push.
% Early notifications as such can aid developers
% from introducing a dependency with already known vulnerabilities.
% However, additional static and dynamic analysis
% to assess the risk of the involved vulnerability
% can take longer time as observed in the case of Steady. 
% A single project, OpenMRS-Core, took 3 hours and 9 minutes
% to complete the two additional analysis for Steady.

\begin{tcolorbox}
    We find SCA tools to vary widely in the reporting of known vulnerabilities, for both Maven and npm dependencies. 
\end{tcolorbox}

\subsection{Why do the tools differ on vulnerability reporting?}\label{sec:manual}
We list the reasons we characterized
(with no particular order)
through manual analysis behind the differences in the tools' results:

\begin{enumerate}
%1. Detecting dependencies outside package manager, 
% possibly through source code scanning
    \item \textbf{OWASP DC and Com. C detected 
JavaScript\\dependencies in Maven projects:}
The Maven projects in OpenMRS can also contain front-end JavaScript files. OWASP DC was able to identify dependencies from JavaScript files such as \textit{jquery}, \textit{handlebars}. These dependencies are not resolved by Maven itself or declared in any dependency manifest file. Besides OWASP DC, only Com. C detected Java\-Script dependencies in Maven projects. In total, 42 Java\-Script dependencies were found by OWASP DC, while Com. C found 20.

% 2. are internal dependencies relevant for SCA tools?
\item \textbf{Only OWASP DC reported vulnerabilities in internal dependencies:} 
As mentioned in Section \ref{sec:openmrsdeps}, OpenMRS can have internal dependencies 
% -- one project within OpenMRS distribution listed as a dependency to another project -- 
which were reported only by OWASP DC. OWASP DC reported 200 dependencies which are OpenMRS projects. However, these 200 dependencies contain only 14 CVEs and 6 non-CVEs. OpenMRS projects are divided into many submodules and OWASP DC reports the same vulnerability separately for each submodule, which results in an inflation of reported dependencies. 
% On the contrary, the failure of other tools in recognizing vulnerabilities in OpenMRS projects indicates that incompleteness in a tool's vulnerability database may result in false negatives.

% 3. same vulnerability reported under multiple packages: inaccurate vulnerability to package mapping
\item \textbf{Same vulnerabilities can be repeated over multiple \\packages:} 
We observe tools may report the same vulnerability across many related packages, such as dependent packages of a vulnerable package.
For example, CVE-2014-3625 was only reported for \textit{spring-webmvc} by MSV, Snyk, Steady, and Commercial A. However, OWASP DC reported this CVE for 
five separate \textit{spring} packages
as NVD simply lists the whole \textit{spring-frameowrk} 
as affected by this CVE. 
% Similarly, we find that CVE-2014-0114 for the package  \textit{commons-beanutils} is also listed under 3 \textit{struts} packages by OWASP DC and under \textit{struts-core} by Commercial A. \textit{commons-beanutils} is a dependency of \textit{struts} package, and therefore, CVE data reports both the products as affected by the CVE
%which may be the reason behind such duplicated reporting. 
Conversely, OWASP DC detected functions of npm packages as individual dependencies. In the \textit{lodash} package, OWASP DC detected 31 functions, such as \textit{lodash.\_baseassign} and \textit{lodash.\_reevaluate}, separately besides the package itself, and repeated the same 7 vulnerabilities for each of them while other tools simply reported the \textit{lodash} package as vulnerable.

Prior work reported that relying on the Common Platform Enumeration (CPE) identifier that comes with CVE data 
may be a reason behind inaccurate vulnerability to package mapping~\cite{cpeveracode}.
For example, OWASP DC reported the same 17 CVEs for 
\textit{activeio-core}, \textit{activemq-core}, and
\textit{kahadb} as they all map to the same CPE identifier
while other tools only reported \textit{activemq-core}.

% 4. inaccurate mapping of affected versions
\item \textbf{Tools may have different mapping of vulnerability to affected versions of packages:}
Incorrect mapping of a vulnerability to the affected version range of a package may result in inaccurate alerts. 
For example, in \textit{commons-beanutils:1.7.0},
OWASP DC, Com. C, and Commercial A reported 
CVE-2014-0114 and CVE-2019-10086
 while MSV reported only CVE-2014-0114
; Dependabot reported only CVE-2019-10086; and Snyk reported no CVEs at all. 
To investigate this difference,
we looked into Snyk's vulnerability database~\cite{snykdb} and found that Snyk lists the affected version range as $[1.8.0,1.9.2)$ 
and $[1.9.2,1.9.4)$ respectively for the two CVEs and therefore, considers the version OpenMRS uses as free of these vulnerabilities. Similarly, Dependabot lists $[1.8.0,1.9.2)$ version range as affected for CVE-2014-0114 but all versions below $1.9.4$ as affected for CVE-2019-10086. In NVD, the affected versions for the two CVEs are listed simply as up to $1.9.1$ and up to $1.9.3$.
% Again, CVE-2014-3576 is reported by Snyk in \textit{activemq-core:5.4.3} but not by Com. C and Commercial A. We see that Snyk database lists $[0,]$ version range as affected by the CVE while Commercial A
% lists $5.8.0-5.10.1$. 

Similarly, in the npm ecosystem, CVE-2018-1000620 was detected by all tools except Snyk for \textit{cryptiles:0.2.2}. We found that Snyk's database lists $[3.1.0,3.1.3) || [4.0.0,4.1.2)$ 
range as affected by this CVE. The NVD CVE data simply lists version up to $4.1.1$ as affected by the CVE.

% 5. (not) considering lock files
\item \textbf{Dependabot reported transitive dependencies through \textit{lock} files:} We notice that Dependabot typically only detect 
direct dependencies. 
In the two cases where Dependabot reported transitive dependencies were due to: a) the Maven dependency file explicitly declared the required version for the transitive dependency, and b) the \textit{lock} file was present in the repository that declared the resolved versions of the full dependency tree. Further, no other tools reported dependencies from \textit{lock} files except Dependabot, which detected 15 vulnerable dependencies from lock files.

% 6. dependencies under use in runtime
\item \textbf{Commercial B only reported vulnerabilities in dependencies under use during runtime:} As Commercial B tracks dependencies through interaction testing as explained in Section \ref{sec:runningtools}, the tool only reported dependencies that were under use by OpenMRS during integration testing, which explains the low count of dependencies reported. 
% From the context of OpenMRS, dependencies reported by Commercial B may be deemed as of higher risk. However, determining which vulnerabilities reported by the SCA tools actually pose a risk to the application is out of scope for this study. 

% 7. state of CVEs - not yet included, rejected?
\item \textbf{The state of CVEs may result in differences in tools' results:}
After a CVE is published, the CVE may become \textit{reserved}, \textit{disputed}, or \textit{rejected} based on new information~\cite{cvestate}. We observed that the state of CVEs may be one possible reason behind differences in CVE reporting, as SCA tools need to be timely updated and verify the changes in CVE states. For example, CVE-2019-10768 and CVE-2020-7676 in npm projects were detected by Snyk, Dependabot, and Com. C, but not by OWASP DC and npm audit. However, the latter two tools reported one of them as non-CVEs with a more elaborate explanation. The other CVE is \textit{awaiting reanalysis} (subject to further changes) which may be a possible reason they are not incorporated by the latter tools. Further, we found four rejected CVEs to be reported by Com. C and Snyk, which were not reported by other tools.

% 8. unique non-CVEs
\item \textbf{Tools can report unique non-CVEs not reported by other tools:} 
A comparison between non-CVEs across different tools requires manual analysis, as there is no common identifiers. We manually looked at a random sample of dependencies that were reported to have non-CVEs by multiple tools, and found that each tool reported non-CVEs that
were not reported by any other tools in the study set.

For example, we observe the following cases in \textit{angular:1.6.1}:
OWASP DC reports two \textit{improper input validation}
vulnerability not reported by any other tool.
While Snyk, Com. C, and Dependabot reported
a similar \textit{XSS} vulnerability, Snyk and Com. C also reported unique \textit{XSS} not reported by others. Snyk also reported a unique \textit{denial of service} not reported by the other tools. npm audit did not report any of these non-CVEs.
We noticed similar differences in non-CVEs for other packages as well, e.g. \textit{lodash}, \textit{ws}.
\end{enumerate}

\begin{tcolorbox}
    We categorize 8 reasons behind differences in vulnerability reporting among the studied SCA tools, such as inconsistency in vulnerability to affected package version mapping. 
\end{tcolorbox}

\subsection{RQ2: What metrics are presented by the SCA tools to aid in the risk assessment of dependency vulnerabilities?}\label{sec:rq2} When a vulnerability lies in a dependency, the risk of the vulnerability may need to be determined by how the application uses the dependency -- that is -- in the context of the dependency. We have observed that the studied SCA tools reported several metrics in scan reports to aid in such contextual assessment. We characterize these metrics into five categories:

% When reporting a VD, 
% tools report the known vulnerabilities in the dependencies.
% While all CVEs have a CVSS~\cite{cvss} (Common Vulnerability Scoring System)
% rating associated with them,
% most non-CVEs also have a severity rating 
% provided by the tool. 
% However, this severity rating is a characteristic
% of the vulnerability itself, 
% and the rating is measured from the context
% of the package containing the specific vulnerability.

% We notice that,
% no VD detection tool reports any risk rating
% of a vulnerability in dependency
% from the context of the application itself.

% However, the studied VD detection tools reported
% several additional metrics in their scan reports.
% We identified the metrics that may \textit{possibly} 
% aid developers in assessing the risk of the vulnerability
% in their project's dependency
% and characterized them in five categories, 
% as discussed in the next five subsections:

\subsubsection{\textbf{Code analysis-based metrics}} \label{sec:codebased}
Tools may analyze source code or binaries to infer dependency usage and vulnerability reachability. Three of the tools, Steady, Commercial A, and B, use code analysis-based metrics for Java language:
\begin{table*}[]
    \centering
    \begin{tabular}{lrrr}
         \multicolumn{4}{c}{\makecell[c]{\textbf{Steady: Static Analysis (Vulnerable code potentially executable})}}  \\
         \hline
        \makecell[l]{Total Alerts} & \makecell[r]{Package not in use} & \makecell[r]{Non-vulnerable code of package used} & \makecell[r]{Vulnerable code of package used} \\
        2,489 & 2,095 (84.2\%) & 340 (13.7\%) & 54 (2.1\%) \\
        %\hline
        \multicolumn{4}{c}{\makecell[c]{\textbf{Steady: Dynamic Analysis (Vulnerable code actually executed})}}  \\
        \hline
        \makecell[l]{Total Alerts} & \makecell[r]{Package not in use} & \makecell[r]{Non-vulnerable code of package used} & \makecell[r]{Vulnerable code of package used} \\
        2,489 & 2,437 (97.9\%) & 11 (0.4\%) & 41 (1.6\%) \\
        %\hline
        \multicolumn{4}{c}{\textbf{Commercial A: Vulnerable call chains}} \\
        \hline
        Total Alerts & Vulnerable Method Calls & Total Vulnerable Call Chain & Median Call Chain per Method \\  
        2,998 & 31 & 93 & 2.0\\
    \end{tabular}
    \caption{Code analysis based prioritization metrics: Vulnerable code reachability analysis}
    \label{tab:codemetrics}
\end{table*}

\textbf{Reachability Analysis:} Tool can curate their
vulnerability database with details on which 
part of the code (e.g., method, class) is involved in
a specific vulnerability. 
Tools then can infer if the vulnerable code is reachable from the dependent application through static and/or dynamic analysis. Steady and Commercial A provides reachability analysis for each vulnerability in dependency.

Steady constructs static call graphs of an application to infer reachability, referred to as \textit{\textbf{potentially executable}} (static analysis). Steady also looks at the executability traces through unit testing to determine if the vulnerable code  is \textit{\textbf{actually executed}} (dynamic analysis). Commercial A, similarly perform static analysis to identify vulnerable call chains -- that is -- the call chain from the application code that reaches the vulnerable method of the dependency. 
% Both the tools therefore, inform developers on which vulnerabilities in dependency are reachable from the application. 

Table \ref{tab:codemetrics} shows the reachability analysis from Steady and Commercial A. 
We find that for 84.2\% of the alerts, Steady did not find the corresponding dependency to be used by the dependent application. Further, Steady found only 2.1\% of the alerts were \textit{potentially executable} and 1.6\% of the alerts were \textit{actually executed}. However, we found a disconnect between the findings of static and dynamic analysis. Only for 13 alerts, both static and dynamic analysis found the vulnerable code to be in use.
Also, for 11 alerts where dynamic analysis found the vulnerable code to be actually executed, static analysis did not find any part of the dependency containing the vulnerability to be in use at all. This observation may indicate limitations to reachability analysis. Similar to Steady, Commercial A also found a low number of cases where the vulnerable code of dependency can actually be reached from application source code specifying 93 distinct call chains. 
% from OpenMRS application to a vulnerability in dependency. 
% Further, Commercial A monitors only the dependencies under use and therefore, reports the vulnerability under use ....

Static analysis, such as call graph construction for Java, is known to have limitations~\cite{sui2020recall}. The effectiveness of dynamic analysis, such as Steady's, is also dependent on having a good test-suite and test coverage.
% Further, OpenMRS uses mocking for testing in many cases, which Steady skips.
% Additionally, we set Steady to skip whenever faced with a test case failure. 
We see that OpenMRS projects reach only around ~20\% test coverage in Steady.
The limited test coverage may have affected Steady's findings.
% Nevertheless, the additional information on reachability for certain alerts
% may aid developers in assessing the risk 
% and how the corresponding
% vulnerability can be exploited from the context of their application. 

\textbf{Dependency Usage:} 
The client application may only use a subset of the functionalities offered by a dependency. The code proportion of a dependency used by an application may indicate the probability of a vulnerability being reachable. Steady and Commercial B reports how many classes out of the total available are used in a dependency. For example, Commercial B found 203 out of 414 classes (49\%) for \textit{spring-web} and 790 out of 4,414 (17.9\%) classes for \textit{groovy-all} to have been used by OpenMRS.

\subsubsection{\textbf{Package Based metrics:}}
The characteristics of the dependency package itself
may indicate the risk associated with it.

\textbf{Package security rating: }
Commercial B provides a letter grade on their assessment of the security of a package. The tool calculates the security rating of a package
based on its age, count of released versions, and number of known vulnerabilities. Out of the 17 packages being identified as vulnerable by Commercial B, 16 have an \textit{F} rating, while one has a \textit{D} rating.

\subsubsection{\textbf{Dependency characteristics based metrics}:}
The scope and depth of the dependency
may indicate the risk of the vulnerability
it contains in the context of the application.

\textbf{Dependency scope:} 
For Maven projects, only Steady reported the scope for each dependency. For npm projects, Snyk and npm audit mentions the dependency scope. 

\textbf{Dependency depth:} Risk may be associated with how deep a dependency lies within the dependency tree. Only Snyk and Steady indicate if a dependency is direct or transitive for Maven projects. For npm, Only Snyk and npm audit reports all possible dependency paths for each vulnerability, indicating the possible depths.

\subsubsection{\textbf{Vulnerability based metric:}}
The characteristics of the vulnerability itself can be used in assessing risk. 
We found three types of information provided by the tools:

\textbf{Severity:} 
The industry standard for rating the severity of vulnerability is the Common Vulnerability Scoring System~\cite{cvss} (CVSS), which are publicly available for CVEs. For the non-CVEs, Snyk and Commercial A also present a CVSS score. However, Dependabot and npm audit present a severity rating on a scale of their own for both CVEs and non-CVEs. For both the tools, the scale consists of four levels similar to CVSS3 levels: \textit{low}, \textit{moderate}, \textit{high}, and \textit{critical}

% We compared these two tools' rating with CVSS3 scores for the CVEs they reported and found that both tools 
% generally have a higher rating than CVSS3 for the same CVEs. 
% npm audit had a higher rating than CVSS3 for 15 of the 30 CVEs (50\%) that it detected (1 CVE does not have a CVSS3 score).
% % -1 2
% % 0 13
% % 1 13
% % 2 1
% % 3 1
% Similarly, Dependabot had a higher rating for 54 of the 86 CVEs (67.5\%) that it detected (4 CVEs did not have a CVSS3 score).
% -1 2
% 0 30
% 1 42
% 2 12

\textbf{Available exploits:} 
The availability of known exploits may contribute in assessing the risk for a vulnerability in a dependency. For each vulnerability, Snyk provides information on whether an exploit is publicly available. For the 310 Snyk vulnerabilities, Snyk reports that 218 do not have a public exploit; 10 have a functional exploit; 37 have a proof of concept exploit; while 45 have unproven exploits available. Commercial A also reports on available exploits.

% \textbf{Code location:} The exact code location of the vulnerability can help developers to identify if their application uses the corresponding dependency feature.
% Snyk and Steady provides the functions that were changed 
% in the 
% patch that fixed a vulnerability while Commercial A also provides the patch link
% if available.

\textbf{Popularity:} How popular or well-known is a vulnerability may indicate the probability it may get exploited in the wild. Steady integrates Google trend analysis for each vulnerability in its reports, which indicates the count of search hits within the past 30 days for the CVE identifier.

\subsubsection{\textbf{Confidence in alert validity:}}
SCA tools may provide a confidence rating for each alert, which may aid developers in prioritizing auditing.

\textbf{Evidence count:} As OWASP DC detects dependencies from scanning multiple sources, it can provide an evidence count as a proof for each dependency. The tool provides a confidence label, from \textit{Low} to \textit{Highest}, based on this evidence count. For all alerts except 4 Maven alerts, OWASP DC provided with either \textit{High} or \textit{Highest} confidence rating. 
% Similarly, all npm alerts were labeled with 
%  \textit{Highest} confidence.
\begin{tcolorbox}
    Our tool survey highlights the research need for a
    common risk measurement framework specifically for dependency vulnerabilities that can be adopted by the SCA tools. 
\end{tcolorbox}

\section{Discussion}
Below, we discuss observations and implications from our work:

% \todo{Recommendations on how to build a ground truth: Lessons learned }

% \textbf{Future research is required to answer \textit{what makes a vulnerability in dependency false positive.}} Prior research has shown that not all vulnerabilities in dependency may be relevant for the application~\cite{pashchenko2020vuln4real,pashchenko20qualitative}. In our study, we find Commercial B to find less number of vulnerable dependency to be under use during integration testing. We also find Steady and Commercial A reporting low number of vulnerabilities in dependency to be actually reachable from the application. However, we have shown that there can be limitation to such reachability analysis. Further, how do developers and security testers assess the risk of vulnerable dependencies has not been studied yet. Future research opportunity lies in establishing and evaluating frameworks and metrics to measure the risk of vulnerabilities in dependencies.

\textbf{Resolving dependencies only through dependency manifest files may not provide all the open source code in use.} We find that OWASP DC and Com. C detect JavaScript dependencies in Maven projects through source code scanning,
which are not declared in the Maven dependency file. This observation sheds light on the importance of scanning source code, binaries, and deployment environments to resolve all the open source code used by an application. Further, projects can use code fragments from open source packages~\cite{scaevaluation} that should also be identified in order to report known vulnerabilities. Future research is necessary to understand how reliably we can identify all the open source components used in a deployed application.

\textbf{Accurate mapping of vulnerability to affected versions of packages should be ensured by the tools to avoid false positives:} 
We showed examples in Section \ref{sec:manual} on how inconsistency in vulnerability to package mapping can result in differences between tools' results. We also find inconsistencies in what version range is listed as affected for a specific CVE by different tools. Our findings highlight the importance of maintaining the accuracy of the tools' vulnerability databases. 

\textbf{Non-CVEs should get reported to CVE database to establish their validity and cross-tool vulnerability mapping.} We find that SCA tools report known vulnerabilities in dependencies that do not have a CVE identifier. To understand why the non-CVEs might not have been incorporated into the CVE database, we look at their publication date. Of the 53 non-CVEs reported by Snyk, 41 were published before 2020; while of the 54 non-CVEs reported by Com. C, 50 were published before 2020. Therefore, developers may question why a reported vulnerability does not have a CVE identifier, as CVE validation usually takes around three months. Furthermore, we see that different tools can report the same non-CVEs. However, cross-tool referencing cannot be done automatically without a common identifier like CVE. We suggest reporting the non-CVEs to CVE database to establish validity, make cross-tool referencing easy, and make the vulnerability widely known.

\textbf{Developers may employ more than one tool to leverage different vulnerability databases:} The reporting of unique non-CVEs by the tools highlights the existence of known vulnerabilities in the wild, not necessarily tracked by a centralized database. SCA tools may have delay in incorporating such non-CVEs, if not completely missing them.
Prior work has shown that unawareness of the vulnerability is a major reason developers do not update vulnerable dependencies~\cite{kula2018developers}. Therefore, our findings suggest developers may use multiple SCA tools to be timely informed of all the known vulnerabilities.

\textbf{Tools should suggest fix options while explaining the risk of any potential backward incompatibility.} 
Out of the studied tools, Snyk, Dependabot, npm audit, and Com. C offer automated fix suggestions.
% In the case of npm projects, npm audit suggests a fix option for each vulnerable dependency path. However, for Maven projects, the fix of the transitive dependencies may require manual intervention to analyze the potential risk of a version change. 
However, prior work has found that fear of breaking change is one of the primary reasons developers do not want to update vulnerable dependencies~\cite{patchingfear}. Tools may provide a more in-depth analysis on what code change is there with a certain version change in a dependency and if there are any possibilities of introduction of regression bugs. 

\textbf{Developers may need to evaluate security risk of dependency vulnerabilities case-by-case basis:} 
Prior research has shown that not all vulnerabilities in dependency may be relevant for the client application~\cite{pashchenko2020vuln4real,pashchenko20qualitative}.
We find two tools to offer reachability analysis for each dependency vulnerability. 
However, future research is required to evaluate if developers in the real world actually use such metrics and 
find them useful or not.
The lack of a common framework for risk assessment 
among SCA tools
suggest that developers will have to evaluate the vulnerability alerts case-by-case basis based on their expertise on the project codebase and how the dependencies are being used
by the specific project.

% \textbf{Future research is required to evaluate metrics that can aid in assessing the risk of a vulnerable dependency.} 
% We find that no tools provide any aggregated risk rating of a vulnerability in dependency from the context of the dependent application. Prior work has found that many vulnerabilities in the dependency may not be relevant to the application itself~\cite{zapata2018towards}. 

% \subsection{Effect of project Structure}
% Only Snyk can find all dependency file from the root directory autmoatically. Maven plugins can find Maven submodules. But tools like npm audit, OWASP has to be fed each dependency file directly. So a multi-project structure may confuse some of the tool.

% \subsection{Continuous vulnerability discovery}

% How many vulnerability since release? Give a timeline map.

% \item npm audit can have same CVE listed under more than one vulnerabilities in it's own database based on who reported and how it is described. For e.g., 	"CVE-2019-10744" is listed under both id 1523 and 1065.
% Github found dependencies from old package-lock.json file that npm no longer lets us to working with `Eintegrity' error. So when we remove lock file and do npm install again we don't find this dependency..

\section{Limitations}
We evaluate 9 SCA tools on one web application at a certain release point, which poses a threat to the generalizability of our findings. However, OpenMRS consists of 44 projects with 547 Maven and 2,213 npm dependencies, making our case study suitable for a comparative evaluation. Further, the single case study enables us to look in-depth with manual analysis on why the tools' results differed. 
% The observations we listed in Section \ref{sec:manual} explains how the studied SCA tools will differ on other Maven and npm projects as well. 
However, our findings may not generalize to other package ecosystems, such as Ruby or Rust. Similarly, any other application with a difference in dependency management may have yield different findings. For example, the application that actively maintains its dependencies updated will have a few vulnerable dependencies overall and therefore, would not show large differences among tools' results. However, that would not invalidate the differences we observed for the studied SCA tools on OpenMRS. Another threat to the external validity involves the selection of the SCA tools. We explain our decision criteria in Section \ref{sec:selection}. 
% We stopped including new tools for this study when we were unable to find any unique new functionalities from the three dimensions mentioned in \ref{sec:selection}. 
While we are unable to cover all existing SCA tools, we do not claim the findings we have in Section ~\ref{sec:manual} and ~\ref{sec:rq2} to be exhaustive.

Another limitation of our study involves the absence of ground truth. In the context of SCA alerts, there can be three steps for building such a ground truth - a) determining all the open source dependencies in use; b) determining the correctness of the reported vulnerability data; c) determining exploitability of the dependency vulnerabilities. 
These steps are nontrivial to perform manually
for a real software 
and may be more suitable with a synthetic test subject~\cite{delaitre2018sate}. Regarding exploitability, 
no tools except Commercial B claims to filter out dependencies on any use criteria, while Steady and Commercial A only offers additional analysis to aid in such contextual assessment.
Therefore, exploitability would not be a fair comparison criteria 
for the studied SCA tools,
and out-of-scope for this study. 
Similarly, how developers respond to alerts from SCA tools is also not in the scope of this study. 
Further, for RQ1, we do not conduct any statistical test to measure significance in difference, as we only perform a single case study.

\section{Related Work}\label{sec:relwork}
The dependency network of package ecosystems and the presence of known vulnerabilities in dependencies have been studied in the literature~\cite{decan2019empirical,kikas2017structure}.
Decan et al.~\cite{decan2018impact} studied the impact of security vulnerabilities in npm dependency network and found that the number of packages with a known security vulnerability is growing over time, and half of the dependent packages do not get fixed even when the fix is available. \cite{hejderup2015dependencies} and ~\cite{lauinger2018thou} also found around one-third of the packages in the npm network to have at least one vulnerable dependency, while \cite{hejderup2015dependencies} found that \textit{context use of the module} and \textit{breaking changes} are potential reasons for not resolving these dependencies.
The potential impact of the vulnerabilities in dependency has also been studied. Zapata et al.~\cite{zapata2018towards} studied the impact of a vulnerability in \textit{ws} package in npm network on applications that were using the vulnerable version of the package. The study finds 73.3\% of the dependent applications did not actually use the vulnerable code.  The study also finds that the dependent applications that do not use the vulnerable code take longer to migrate to new versions of a dependency. To detect dependencies where the vulnerable code is 
actually used by the including application, Ponta et al.~\cite{plate2015impact,ponta2018beyond,ponta2020detection} proposed a \textit{code-centric} and \textit{usage-based} approach based on which the tool Steady is developed. Paschenko et al.~\cite{pashchenko2020vuln4real} discuss
the over inflation problem when reporting dependencies with unexploitable vulnerabilities. In another work, Paschenko et al.~\cite{pashchenko20qualitative}
interviewed developers on dependency management. The study found that developers think SCA tools generate many \textit{irrelevant} and \textit{low-priority} alerts, and may even rely on social channels than SCA tools for vulnerability reporting. Developers recommended SCA tools to \textit{report only relevant alerts, work offline, and be easily integrated into company workflow}. The literature focuses on the importance of contextual assessment of vulnerable dependencies.
To the best of our knowledge, there has been no study yet evaluating and comparing the existing SCA tools. The closest to our work is a recent study by Ponta et al.~\cite{ponta2020detection} where the authors compared their research tool Steady with OWASP Dependency-Check. The study compares the two tools over a sample of alerts generated on Java applications. The comparison was performed from the perspective of the reachability of a vulnerability in the dependency. The study finds both tools to have their unique findings. The study also finds Steady to have no false positives but a few false negatives, while OWASP DC has non-negligible false positives. However, the study focuses on evaluating the detection capabilities of Steady based on its reachability analysis, whereas our study aims to provide a comprehensive comparison of existing SCA tools for both Java and JavaScript dependencies.
% \begin{itemize}
%     Kuala et al.~\cite{kula2018developers} talks about how developers update dependencies and if security advisory plays any role. Talk about our FSE Paper.
%     \item Evaluation of security tools. How these are done and what are the generic findings?
%     \todo{Prior example of tool evaluation or comparison work?} SATE paper. Recent paper on comparison of technical debt measuring tools.
% i remember a work where the authors investigate VDs reported by owasp dvc and finds most of them are false positives
% \end{itemize}
\section{Conclusion and Future Work}
We evaluate 9 SCA tools on a large web application 
composed of Maven (Java) and npm (JavaScript) projects. 
We find that the tools vary 
across a wide range in the count of reported unique vulnerabilities and 
the dependencies that contain these vulnerabilities. 
Evidences in our findings suggest that
accuracy, up-to-dateness, 
and completeness of vulnerability database 
is the key strength of an SCA tool. 
While building automation technologies 
for continuous monitoring of vulnerability data 
from open source package ecosystems
is a future research need,
developers at the moment may leverage multiple SCA tools
in order to not miss any known vulnerability.  
Further, SCA tools should be able to pick up 
open source components beyond 
what is declared in the dependency manifest files, if any.
% We find results by the tools can both be inflated based on how the data is presented
% (e.g. repeating the same vulnerability across several related packages) and deflated based on the scanning technique (e.g. interaction testing). 
% Evidences in our findings suggest that accurate mapping of vulnerability to affected versions of package is required for a reliable SCA tool. We also see that maintaining an updated vulnerability database accommodating recent vulnerability findings from the broad open source ecosystem can also be strength of an SCA tool.
We find that tools can provide code-analysis based metrics 
to assess the risk of dependency vulnerabilities.
However, the effectiveness of such analysis 
needs to be evaluated and 
how developers in the real world responds 
to vulnerable dependencies should be studied. 
We have also seen tools to provide 
non code-analysis based metrics, 
such as package security rating, 
vulnerability exploits, and popularity.
Prior work has found developers to rely on social channels to assess the risk of newly found vulnerabilities in open source packages~\cite{pashchenko20qualitative}. 
As developers may get overwhelmed with 
frequent SCA alerts
and patching may require extensive regression testing~\cite{patchingchallenges}, 
future study is required on 
what metrics and information we can provide developers with
to aid them in assessing and prioritizing 
the fix of vulnerable dependencies. 
% Some areas of future work include how do practitioners prioritize vulnerable dependencies; what metrics they can use in their decision making process; and how effective are those metrics in the real world.

% Overall, we find that tools should strive to maintain an accurate vulnerability database while future research is necessary in evaluating the metrics for assessing the risk of vulnerabilities in dependency.

\section{Acknowledgements}
We thank the RealSearch group and anonymous reviewers
for their feedback.
Our research was funded by NSA.

\bibliographystyle{plain}
\interlinepenalty=10000
\bibliography{bibliography}

\end{document}